\documentclass{article}
\usepackage{afterpage, amsmath, array, booktabs, color, enumerate, graphicx,epstopdf}
\usepackage[margin=1in]{geometry}

\title{Emergence of Distributed Coordination in the Kolkata Paise Restaurant Problem with Finite Information}
\author{Diptesh Ghosh\thanks{Production \& Quantitative Methods Area, IIM Ahmedabad, Gujarat 380015, INDIA. Email: diptesh@iima.ac.in}
\and 
Anindya S.~Chakrabarti\thanks{Economics Area, IIM Ahmedabad, Gujarat 380015, INDIA. Email: anindyac@iima.ac.in; Corresponding author}}
\date{\today}

\begin{document}
	\maketitle

\begin{abstract}
In this paper, we study a large-scale distributed coordination problem and propose efficient adaptive strategies to solve the problem. The basic problem is to allocate finite number of resources to individual agents such that there is as little congestion as possible and the fraction of unutilized resources is reduced as far as possible.
In the absence of a central planner and global information, agents can employ adaptive strategies that uses only a finite knowledge about the competitors. In this paper, we show that a combination of finite information sets and reinforcement learning can increase the utilization rate of resources substantially.
\end{abstract}


\vskip .5cm

%{\bf PACS code:} *******

\section{Introduction}
\label{sec:intro}
Large-scale coordination problems abound in the modern economy and society. Traffic flows, multi-processor computing, matching order flows in financial markets, and supply chains are some situations where coordination is required among agents for smooth functioning of the corresponding systems. In a centralized system like a centralized market, one can think of a central planner or an algorithm to solve the problem of coordination and ensure that resources are allocated to minimize wastage. However, many economic and social systems are characterized by a large number of independently participating agents and a finite number of resources for which the agents compete \cite{Marsili2009}. In such cases, it is prohibitively costly to collect information about all agents to use in centralized planning procedures. Therefore an important problem is to find efficient strategies for individual agents that use minimal computational power, and solve the global coordination problem when used locally by all agents \cite{Chakraborti2015}.

Such autonomous multi-agent systems have a number of distinctive characteristics. Ref.~\cite{Lerman_02} summarizes such properties as follows. The number of participating agents is large and they do not explicitly communicate with each other. The aggregate behavior of the system is robust to individual agent failures, i.e. no single agent's mistake can cause a system-wide disruption. This indicates that the system's ability to adapt as well as it self-organization ability are fundamental features. Economic and social systems are often exhibits close similarities \cite{SenChakrabarti_14}, where both systems evolve close to the optimal state without any global coordinator, occasionally perturbed by exogenous as well as endogenous disturbances.

The Kolkata Paise Restaurant (KPR hereafter) problem was proposed as a generalized multi-agent, multi-choice problem \cite{chakrabarti2009,Ghosh2010} that essentially characterizes such a scenario. There are $N$ agents (persons) who are competing for $N$ resources (restaurants). A resource can serve only one agents at every point in time. The decision problem for an agent is to choose a resource that is chosen by few or none of other agents. While choosing a resource, an agent does not know which resources the other agents are planning to opt for. Thus it is a generalized {\it minority game} \cite{Arthur1994a,Challet2004,Fogel1999} as well.

In the KPR context, a customer's strategy for choosing restaurants in any period is represented by a $1 \times N$ vector of probabilities with which she is going to visit each of the restaurants in that period. The sum of the probability values equals 1. A candidate's probability revision protocol involves redistributing the probability sum among the components of the vector. A candidate is said to be {\it stable} if her probability vector has 0's in all positions except one and 1 in the remaining position. Such a probability vector is called stabilized. A stable customer will visit the same restaurant in all periods unless she revises her probability vector.

Recent research has focused on finding strategies that the customers can employ locally (i.e., strategies that do not use global information) and dynamically to reach a state where the rate of utilization of restaurants is maximized. Several such strategies have been reported. Current literature shows that attaining a utilization ratio of 0.8 (i.e.\ 80\% of the customers are served by 80\% of the restaurants every period in the long run) is achievable \cite{Chakraborti2015}. In this paper, we propose a set of revision protocols that combine local information with reinforcement learning and show that there is dramatic improvement in the utilization rate compared to the strategies proposed in the literature.

There are two important ingredients of the update protocols that we propose. Ref. \cite{Chak_Gho_JEIC}, showed that reinforcement learning is very useful in solving the coordination game. Essentially, it depends on the Pavlovian strategy of {\it win-stay, lose-shift} \cite{chakrabarti2009}. However, mere dependence on such a strategy will require multiple trial-and-error before a high level of utilization rate is achieved. If repeatedly making errors are costly (for example, in case of solving task allocation problems by computer processors or allocating traffic across  roads), then such a strategy is not very efficient. One possible way to enhance faster convergence to the state of high utilization is to allow agents to have bigger information sets. The benefit is that the agents can use information sets to reduce the number of wrong attempts. However, this also comes with a cost. Having a similar information set across multiple agents will force them to opt for the same set of resources and hence, will defeat the final objective of staying in `minority' or avoiding crowd \cite{Sasidevan_16}. Our contribution in this paper is to show that a mix of these two strategies leads to high levels of utilization of resources. A more important result is that such a combination is very efficient as the utilization rates become substantially high within a few iterations.

Ref. \cite{Kalinowski_00} had studied emergence of cooperation in minority games with the help of local information. It was preceded by the study on minority games where agents had access to the history of a random set of agents \cite{Paczuski_99}. Our work is different from those papers in one crucial aspect. The agents considered in such models were Boolean indicating that the choice set of the agents were rather limited. Here on the other hand the number of resources also  scales with the number of agents. Thus the introduction of multi-choice environment makes the solution more computationally intensive. However, we show that there are very simple revision protocols which use local information and are able to solve the coordination problem at the global level.

In this paper we first introduce the KPR problem with reinforcement learning in Sec.~\ref{sec:KPR}. In Sec.~\ref{sec:update} we present six revision protocols for customers that lead to efficient resource utilization. Each protocol has two variants. In Sec.~\ref{sec:results} we simulate the behavior of agents following the revision protocols introduced in  Sec.~\ref{sec:update}. The results from simulations to show that several of these protocols attain utilizations of close to 100\%. We summarize our results in this paper in Sec.~\ref{sec:summary}.

\section{KPR problem with reinforcement learning}
\label{sec:KPR}
Let $I$ be the set of customers ($|I| = N$), $R$ be the set of restaurants ($|R| = N$), and $t = 1, 2, \dots$ be the time periods over which agents and restaurants interact. In the KPR problem, at the beginning of each period, each customer chooses one restaurant which she will visit in that period. Therefore, each restaurant will have zero, one, or more customers choosing the restaurant during any period. If there are no customers for a restaurant, it remains idle for that period. If there is one customer, then the restaurant serves her. If there are more than one customers who choose a restaurant, the restaurant chooses one of them at random and serves her, while the others are not served by any restaurant during that period. Our objective is to determine probability revision protocols that each customer will individually adopt in choosing restaurants so that the fraction of customers being served in a period (which we call utilization fraction) is as high as possible after a number of periods. 
 
In reinforcement learning, initially, each customer assigns a probability $1/N$ of visiting each restaurant in $R$. She then revises her probabilities of visiting various restaurants based on her (possibly limited) knowledge about the allocation of customers to restaurants in the previous time period. An assumption seen in the literature on revision protocols \cite{chakrabarti2009} is that if a customer makes a successful match with one restaurant, then she will opt for that restaurant forever. Such revision protocols have an obvious drawback. Suppose a restaurant serves different customers in different periods. Then those customers will all visit that restaurant in all subsequent periods, and all but one of them will not be served in any of the periods. In the long term, following these revision protocols will result in low utilization fractions. In this paper we propose and experiment with revision protocols in which if a customer has not been served in a particular period, she changes her probability vector in the next period regardless of whether she has been served by any restaurant in the past. This means that in these revision protocols, customers are loyal to a restaurant only as long as the loyalty is reciprocated by the restaurants as well.

\section{Revision protocols}
\label{sec:update}
As mentioned above, in the literature, probability vectors for customers have been revised after assigning customers to restaurants at every period as follows.\vspace{-12pt}
\begin{quotation}\noindent
\begin{enumerate}[Step \mbox{1}.1:]
	\item For each customer served by a restaurant in the current period, revise her probability vector to one in which all entries except the one for the restaurant serving her is 0, and the entry corresponding to the restaurant that served her is 1. (I.e., the customer's probability vector is stabilized.)
	\item For each customer not served by any restaurant in the current period and whose probability vector is not stable, we revise her probability vector.
	\item For each customer not served by any restaurant in the current period and whose probability vector is stable, we do not revise her probability vector.
\end{enumerate}  
\end{quotation}
Different revision protocols differ from one another in the manner by which the probability vector is revised in Step 1.2. For any revising method, if the probability vectors are revised in this way, we call the revise mechanism as Variant 1 of that revising method. 

It has been stated that Variant~1 suffers from the following drawback. Suppose there are two customers $i_1$ and $i_2$ and two restaurants $j_1$ and $j_2$. Suppose both visit restaurant $j_1$ in the first period and the restaurant serves  $i_1$. Then $i_1$ visits restaurant $j_1$ for all subsequent periods. Now suppose in the second period client $i_2$ visits restaurant $j_1$ again. Restaurant $j_1$ can choose customers at random in the perios. In period 2 suppose restaurant $j_1$ serves client $i_2$. Then for all periods after the second one, both $i_1$ and $i_2$ keep visiting $j_1$. Only one of them will be served in each period. However, restaurant $j_2$ remains idle for all periods and the utilization ratio will never exceed 0.5. 

One way to overcome this drawback is to allow all customers who have not been served in a period to revise their probability vectors in the next period. This revision protocol is implemented as follows.\vspace{-12pt}
\begin{quotation}\noindent
	\begin{enumerate}[Step \mbox{2}.1:]
		\item For each customer served by a restaurant in the current period, and was served in the previous period also, we do not revise the probability vector.
		\item For each customer served by a restaurant in the current period, and was not served in the previous period also, we keep a copy of the probability vector for that customer. We then revise her probability vector to one in which all entries except the one for the restaurant serving her is 0, and the entry corresponding to the restaurant that served her is 1.
		\item For each customer not served by any restaurant in the current period but was served in the previous period, we replace her probability vector with the saved copy (refer to Step 2.2). We then revise her probability vector.
		\item For each customer not served by any restaurant either in the current period or the previous period, we revise her probability vector.
	\end{enumerate}  
\end{quotation}
Here too, different revision protocols differ from one another in the manner by which the probability vector is revised in Step 2. For any revising method, if the probability vectors are revised in this way, we call the revise mechanism as Variant 2 of that revising method. 

We now describe different revision protocols for revising probability vectors with limited information. Recall that such revises are made on probability vectors that are not stabilized at the beginning of the period. 

\subsection{Revision Protocol RP1: Local information about competitors}
We assume that the $i$-th customer knows what happened to customers $(i+1)|_N$ through $(i+k)|_N$ in period $t$. Thus every customer knows about a finite subset of customers in a given order. Clearly the information sets of two customers are never identical, although they may intersect.

Suppose customer $i$ visits restaurant $r$ in period $t$ and is not served. Then she revises her probability vector. She resets the probability of visiting restaurant $r$ and all other restaurants that customers $(i+1)|_N$ through $(i+k)|_N$ visited in period $t$ to 0 in the next period and reassigns the probability values that she had assigned to visit these restaurants among the remaining restaurants in proportion to the probabilities that she had assigned them in period $t$. 

Let $V_{it}$ be the set of restaurants that customers $i$ through $(i+k)|_N$ visited in period $t$, and let $P_{it} = \sum_{j \in V_{it}}p_{ijt}$; i.e., $P_{it}$ is the probability that customer $i$ assigned to visiting restaurants that she or one of the customers that she knows about visited in period $t$. If $P_{it} = 1$ then she distributes the probability of visiting restaurants evenly among restaurants not in $V_{it}$ and revises her probability vector to 
\begin{equation}
p_{ij(t+1)} = \left\{\begin{array}{ll}
0 & \text{if }j \in V_{it},\\
1/(N-|V_{it}|) & \text{otherwise.}
\end{array}\right.
\label{eq:NaiveCase1.1}
\end{equation}
If $P_{it} < 1$ then she distributes the probability mass $P_{it}$ proportionally among restaurants not in $V_{it}$. Thus her revised probability vector in this revision protocol is 
\begin{equation}
p_{ij(t+1)} = \left\{\begin{array}{ll}
0 & \text{if }j \in V_{it},\\
p_{ijt}\Big{(}1+ P_{it}{p_{ijt}}/{(1 - P_{it})}\Big{)} & \text{otherwise.}
\end{array}\right.
\label{eq:NaiveCase1.2}
\end{equation}

\subsection{Revision Protocol RP2: Partitioning of information sets} 
In reality people often form clusters of close associates and friends rather than an overlapping string of connections. To model this situation, we assume that the customers are partitioned into groups. Each group knows what happened to other members in the group in period $t$. Suppose that $\mathcal{C}$ is a partition of the set of customers, and customer $i$ belongs to the set $C_i \in \mathcal{C}$ in this partition. We assume that customers in each set of the partition share the results of their previous period's visits with each other. 

The revision protocols of customers in RP2 is very similar to that in RP1. The difference between the two revision protocols is that in RP2 the information at the disposal of each member of a set $C_i \in \mathcal{C}$ are identical, while in RP1, each customer knows the outcomes of the visits of a different and unique set of customers. 

\subsection{Revision Protocol RP3: Partitioning the information sets regarding the resources}
Suppose that customers do not know about other customers but have access to the information about how the restaurants are being utilized. More specifically, restaurants are partitioned into groups. Customer $i$ visiting a restaurant $r$ in period $t$ knows what happened to the other restaurants in the group to which $r$ belonged. Suppose that $\mathcal{R}$ is a partition of the set of restaurants. Customer $i$ who has visited restaurant $r$ in period $t$ knows the status of restaurants in $r \in R^r$ where $R^r \in \mathcal{R}$.

Suppose customer $i$ visiting restaurant $r$ in period $t$ and is not served and revises her probability vector. She reassigns the probabilities that she had assigned to the restaurants in $R_r$ which served customers in period $t$ to restaurants in $R_r$ which did not serve customers in period $t$. She has no information regarding the status of restaurants in $R \setminus R_r$ and hence does not revise her probabilities of visiting them.

Let $V^r_t$ be the subset of restaurants in $R^r$ that served customers  in period $t$, and let $W^r_t = R_r \setminus V^r_t$. Let $P_{it} = \sum_{j \in V^r_t}p_{ijt}$ and $Q_{it} = \sum_{j \in W^r_t}p_{ijt}$. 

If $Q_{it} = 0$, i.e., if customer $r$ was not planning to visit restaurants in $W^r_t$ in period $t$, then customer $i$ revises her probability vector to  
\begin{equation}
p_{ij(t+1)} = \left\{\begin{array}{ll}
0 & \text{if }j \in V^r_t,\\
1/|W^r_t| & \text{if }j \in W^r_t,\\
p_{ijt} & \text{otherwise.}
\end{array}\right.
\label{eq:NaiveCase3.1}
\end{equation}
If $Q_{it} > 0$, then she revises her probability vector to  
\begin{equation}
p_{ij(t+1)} = \left\{\begin{array}{ll}
0 & \text{if }j \in V^r_t,\\
p_{ijt}\Big{(}1+ P_{it} {p_{ijt}}/{Q_{it}}\Big{)} & \text{if }j \in W^r_t,\\
p_{ijt} & \text{otherwise.}
\end{array}\right.
\label{eq:NaiveCase3.2}
\end{equation}

\subsection{Revision Protocol RP4: Imperfect information about the resources} Next suppose that customers are given imperfect information about about the set of restaurants that were idle in period $t$. Suppose the information is that the set of restaurants that were idle in period $t$ was $RI_t \subset R$. Also suppose that the accuracy of the information about idle restaurants was $\alpha$, $0 \leq \alpha \leq 1$. If $\alpha = 0$, the probability that a restaurant chosen at random belongs to $RI_t$ is $1/e$; see \cite{chakrabarti2009}. If $\alpha = 1$ the probability that a restaurant belongs to $RI_t$ is 1 if it was actually idle and 0 otherwise. For $0 < \alpha < 1$ the probability that a restaurant belongs to $RI_t$ is $(1-\alpha)+\alpha/e$ if the restaurant was actually idle, and $\alpha/e$ otherwise.

Now if customer $i$ who was not served in period $t$ believes that the information is correct, i.e., that $RI_t$ is actually the set of idle restaurants, then she would revise her probability vector so that the sum of probability masses assigned to all restaurants in $RI_t$ is 1 as follows. Let $P_t = \sum_{j \in RI_t} p_{ijt}$. If $P_t = 0$ then her probability vector is revised to  
\begin{equation}
p_{ij(t+1)} = \left\{\begin{array}{ll}
1/|RI_t| & \text{if }j \in RI_t,\\
0 & \text{otherwise;}
\end{array}\right.
\label{eq:NaiveCase4.1}
\end{equation}
and if $P_t > 0$ then to 
\begin{equation}
p_{ij(t+1)} = \left\{\begin{array}{ll}
p_{ijt}/P_t & \text{if }j \in RI_t,\\
0 & \text{otherwise.}
\end{array}\right.
\label{eq:NaiveCase4.2}
\end{equation}

\subsection{Revision Protocol RP5: Disbelief about global information}
Now suppose that customers are given correct information about the set of restaurants that were idle in period $t$. However they assign a probability $\pi$ that the information is correct.

Suppose the set of restaurants that were idle in period $t$ was $RI_t \subset R$. If customer $i$ who was not served in period $t$ assigned a probability 1 that this information was correct, then she would revise her probability vector so that the probability mass assigned to all restaurants in $RI_t$ is 1 as follows. Let $P_t = \sum_{j \in RI_t} p_{ijt}$. If $P_t = 0$ then her probability vector is revised to  
\begin{equation}
p^1_{ij(t+1)} = \left\{\begin{array}{ll}
1/|RI_t| & \text{if }j \in RI_t,\\
0 & \text{otherwise;}
\end{array}\right.
\label{eq:NaiveCase5.1}
\end{equation}
and if $P_t > 0$ then to 
\begin{equation}
p^1_{ij(t+1)} = \left\{\begin{array}{ll}
p_{ijt}/P_t & \text{if }j \in RI_t,\\
0 & \text{otherwise.}
\end{array}\right.
\label{eq:NaiveCase5.2}
\end{equation}

In case she assigned a probability 0 that the information was correct, she would disregard this information, and hence revise her probabilities only on the information that restaurant $r$ is not available to her. 

She will revise her probability vector to  
\begin{equation}
p^0_{ij(t+1)} = \left\{\begin{array}{ll}
0 & \text{if } j = r,\\
p_{ijt} \big{(}1 + p_{irt}/(1 - p_{irt})\big{)} & \text{otherwise;}
\end{array}\right.
\label{eq:NaiveCase5.3}
\end{equation}   

Now since customers assign probability $\pi$ to the correctness of the information, customer $i$'s revised probability vector is
\begin{equation}
p_{ij(t+1)} = \pi p^1_{ij(t+1)} + (1-\pi) p^0_{ij(t+1)}\text{ for all }j\in R.
\label{eq:NaiveCase5Combine}
\end{equation}

\subsection{Revision Protocol RP6: Disbelief about imperfect information}
In this protocol, we assume that the information provided about the set of idle restaurants is imperfect, as was the situation in RP4. We also assume that customers assign a probability value between 0 and 1 about the authenticity of the information, as in RP5. This is perhaps the most realistic situation of imperfect information.\bigskip

We next report results from our simulation studies on the revision protocols we have described in this section.

\section{Results from simulations}
\label{sec:results}
In Sec.~\ref{sec:update} we have described six revision protocols of revising customers' probability vectors in the KPR problem. Each of the revision protocols give rise to two variants depending on whether or not customers who were once served by a restaurant remain loyal to that restaurant in all subsequent periods or only up to the time the restaurant serves them. In this section we test the effectiveness of these revision protocols through simulations.

We measure the effectiveness of a revision protocol by its utilization fraction, which is the fraction of restaurants serving customers in a period. The value of the utilization fraction of a good revision protocol comes close to 1 within a few iterations. Incidentally, the utilization fraction for the strategy of choosing a restaurant at random is $1 - 1/e$; see \cite{chakrabarti2009}. Another measure that we use is the stability fraction, which is the fraction of customers who are stable at an iteration. In Variant~1 of each revision protocol, a customer who has become stable at one period remains stable for all subsequent periods. So we expect the stability ratio to increase with increasing iteration number. This is not so in Variant~2, since customers are stable as long as they are being served by a restaurant. Hence in Variant~2, the stability fraction equals the utilization fraction in each period.

\begin{figure}[!tb]	\centering
	\begin{tabular}{m{1.2cm}m{7.2cm}m{7.2cm}}
		\toprule
		Protocol & \multicolumn{1}{c}{Variant 1} & \multicolumn{1}{c}{Variant 2}\\
		\midrule
		RP1 & \includegraphics[width=7cm]{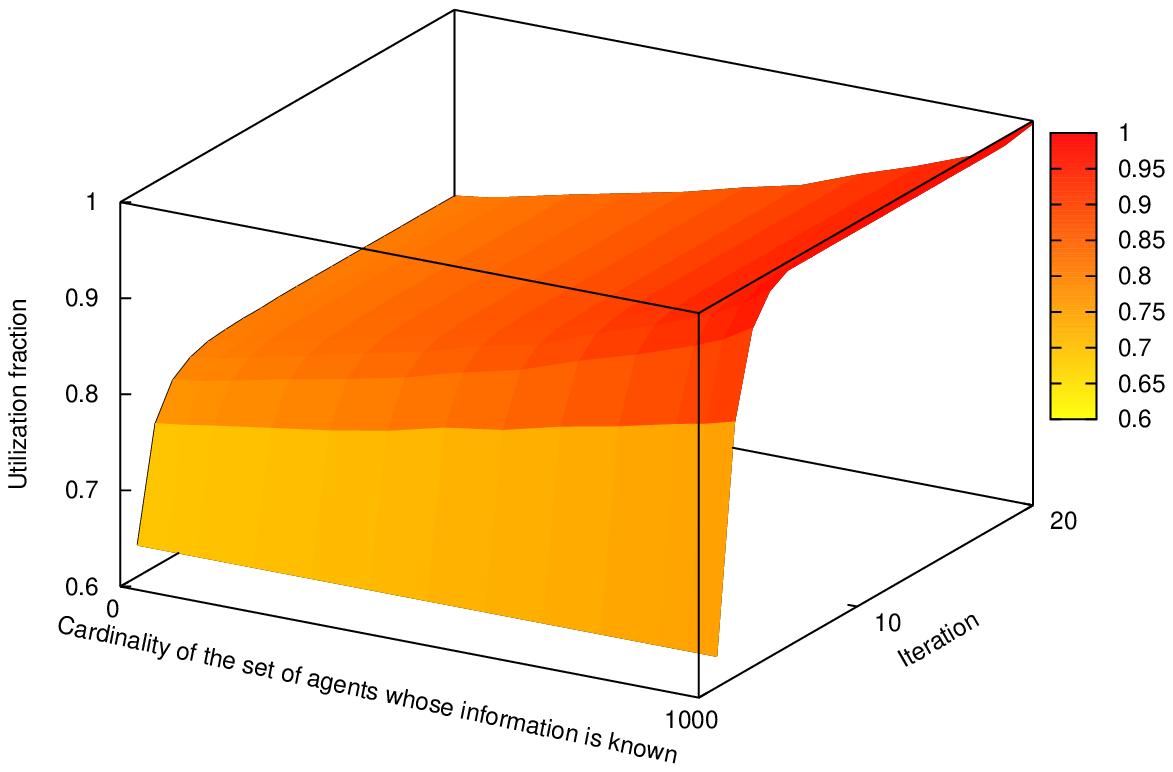} & \includegraphics[width=7cm]{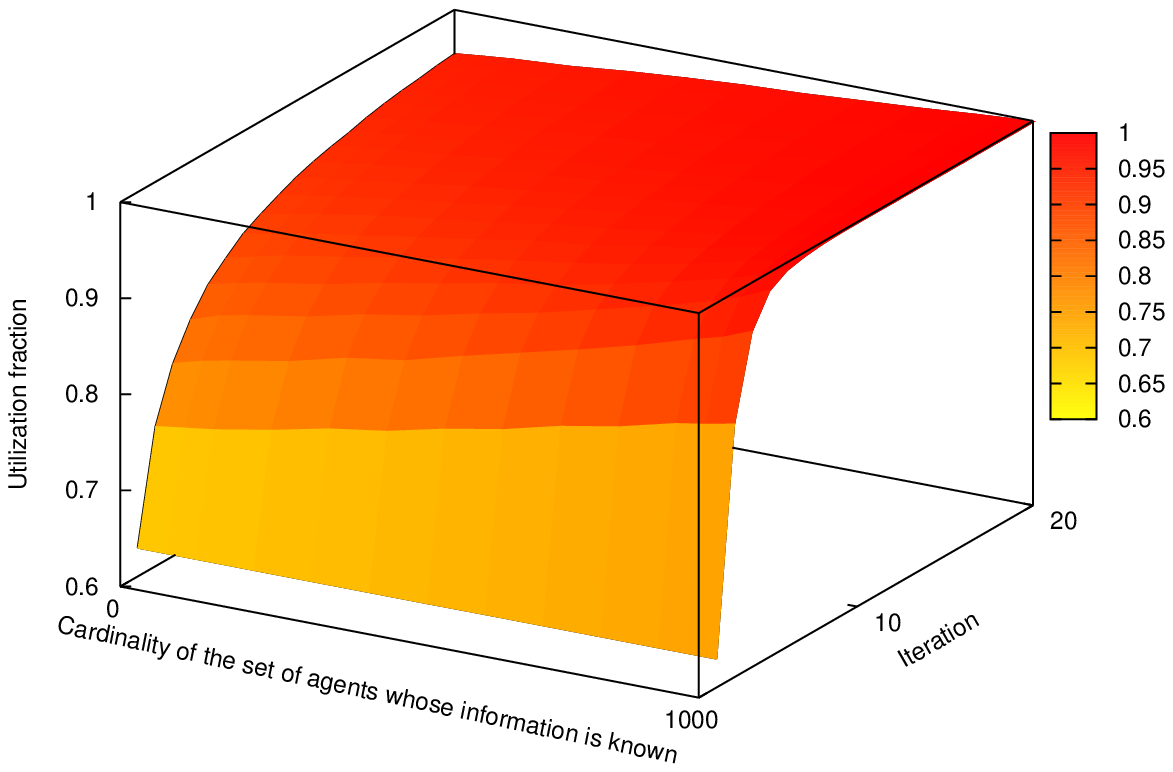}\\
		RP2 & \includegraphics[width=7cm]{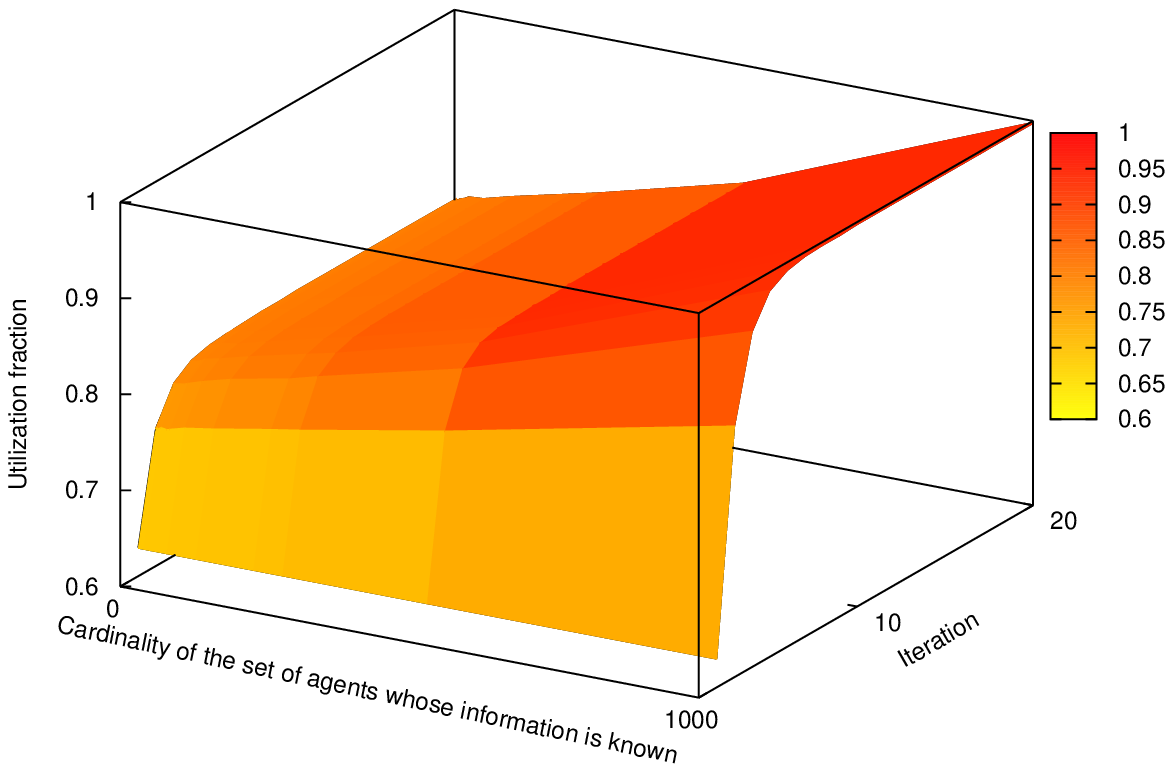} & \includegraphics[width=7cm]{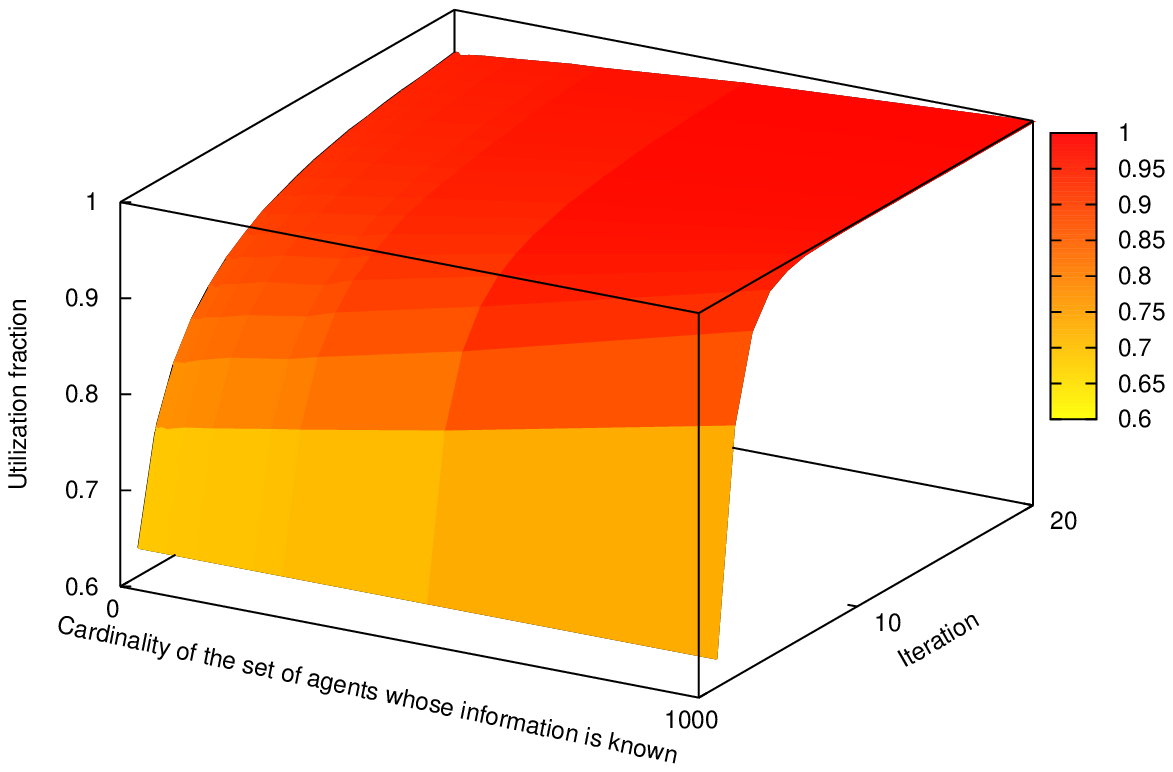}\\
		RP3 & \includegraphics[width=7cm]{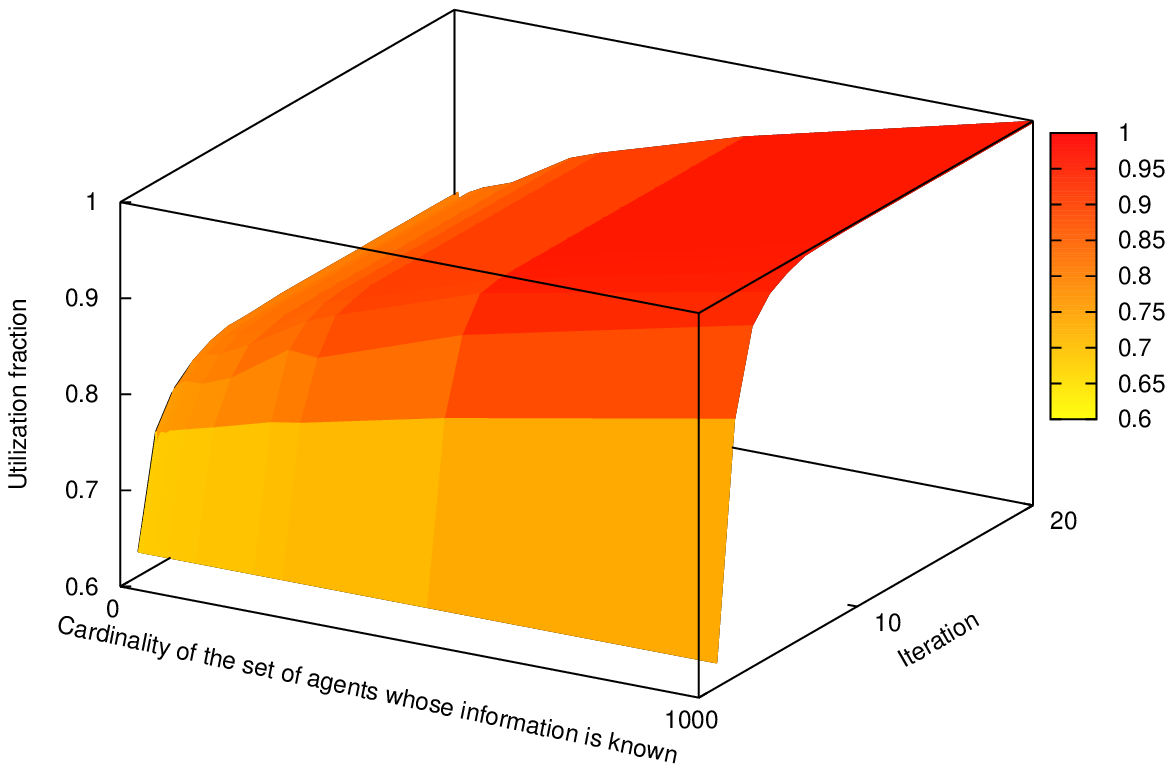} & \includegraphics[width=7cm]{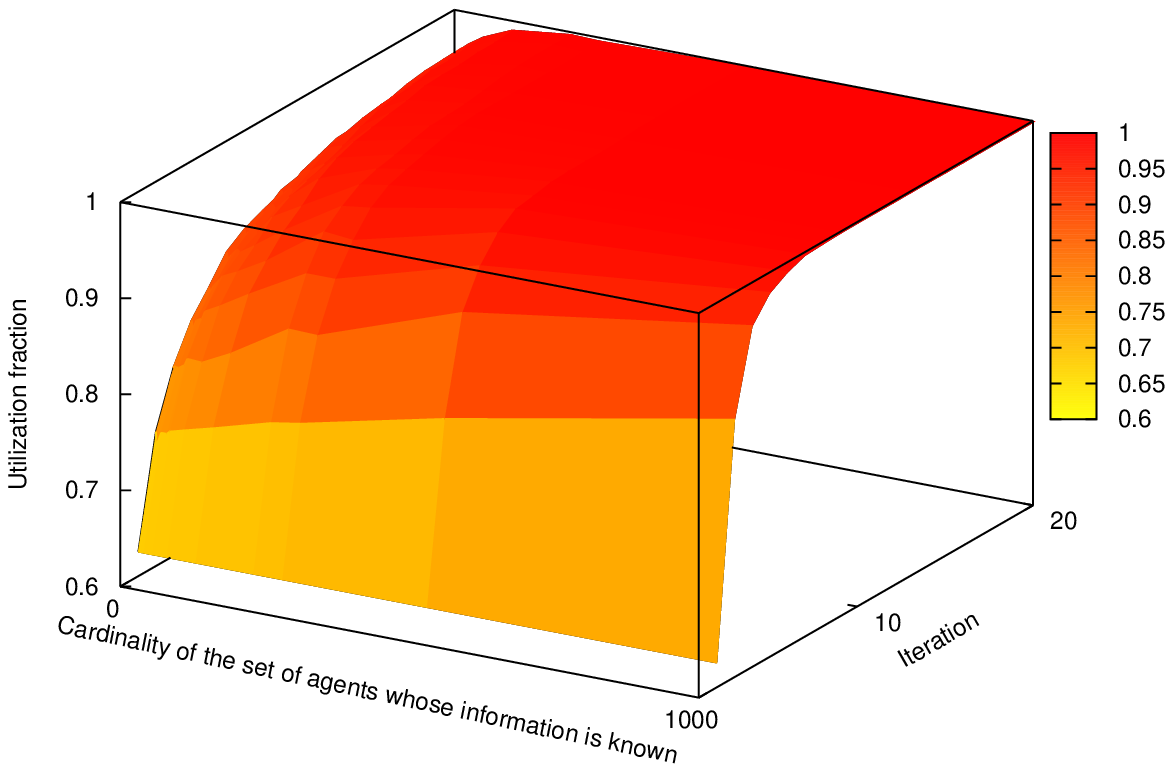}\\
		\bottomrule
	\end{tabular}
	\caption{(Color online) Utilization fractions for various revision protocols.}
	\label{fig:u1}
\end{figure}

\begin{figure}[!tb] \centering
	\begin{tabular}{m{1.2cm}m{7.2cm}m{7.2cm}}
		\toprule
		Protocol & \multicolumn{1}{c}{Variant 1} & \multicolumn{1}{c}{Variant 2}\\
		\midrule	
		RP4 & \includegraphics[width=7cm]{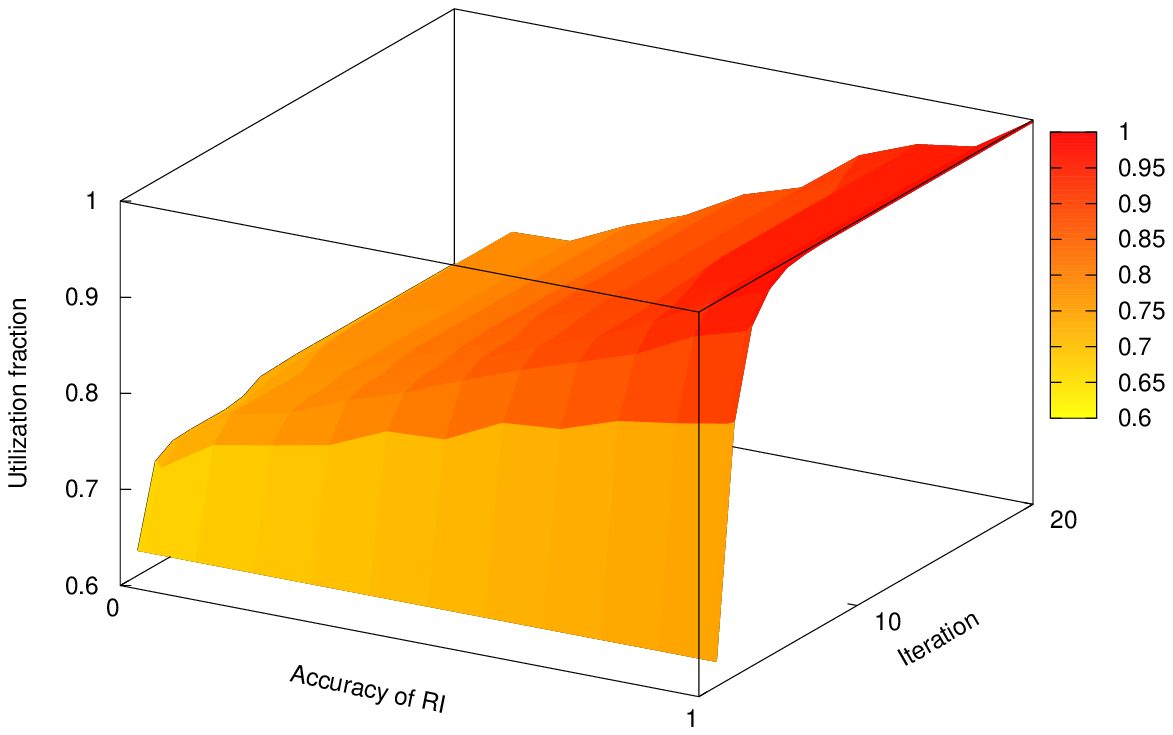} & \includegraphics[width=7cm]{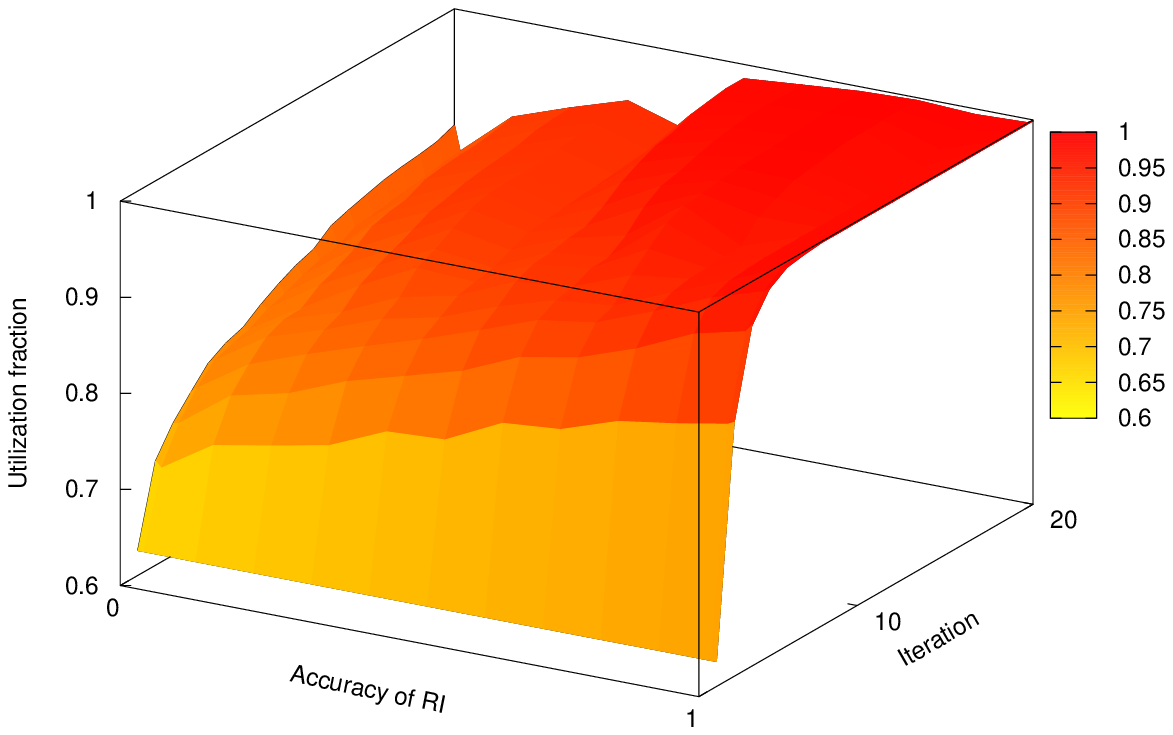}\\
		RP5 & \includegraphics[width=7cm]{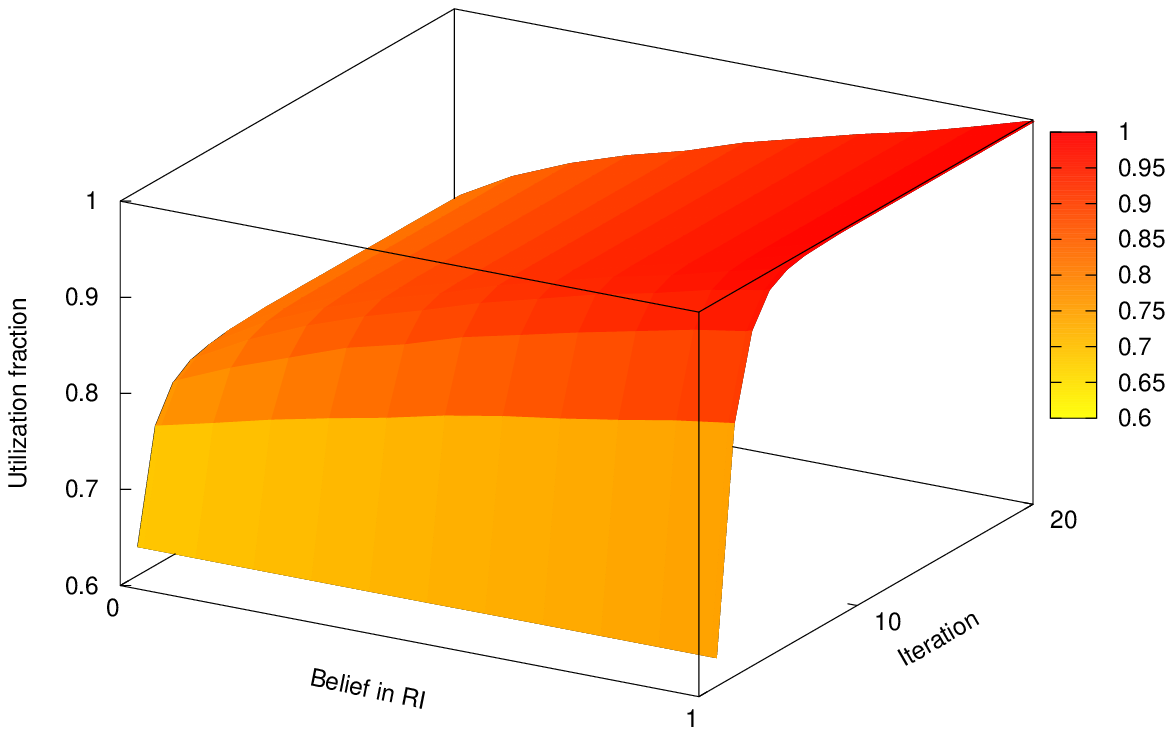} & \includegraphics[width=7cm]{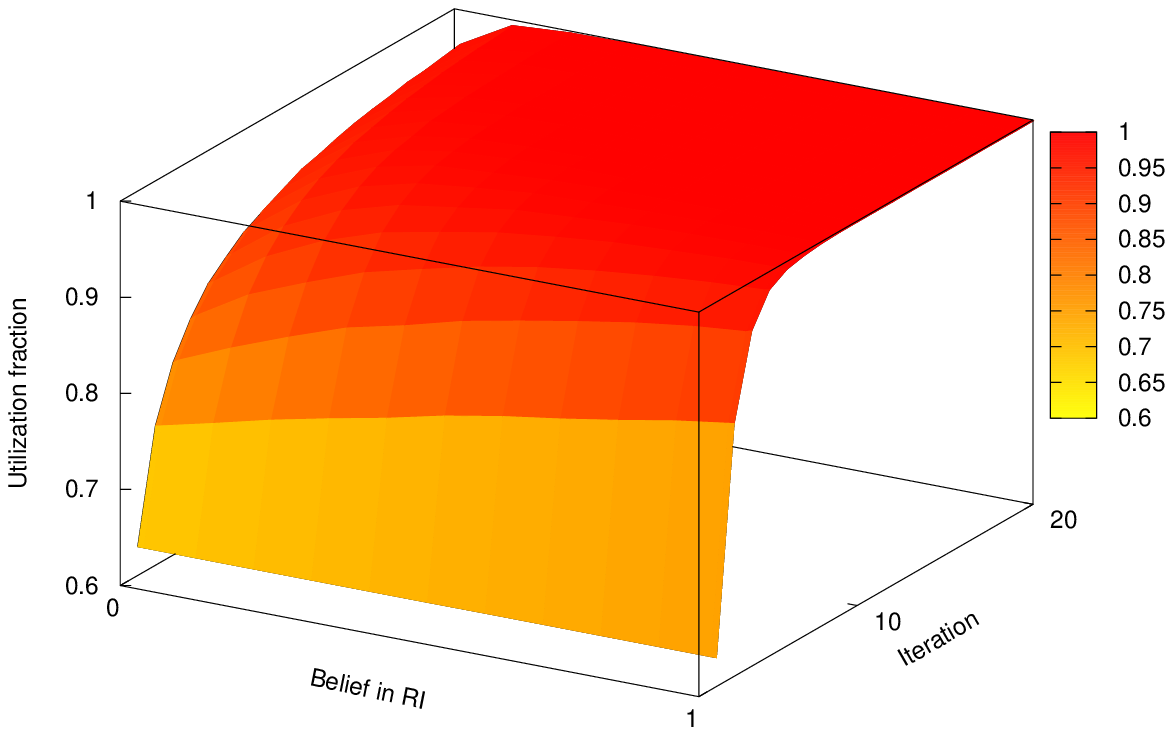}\\
		RP6 & \includegraphics[width=7cm]{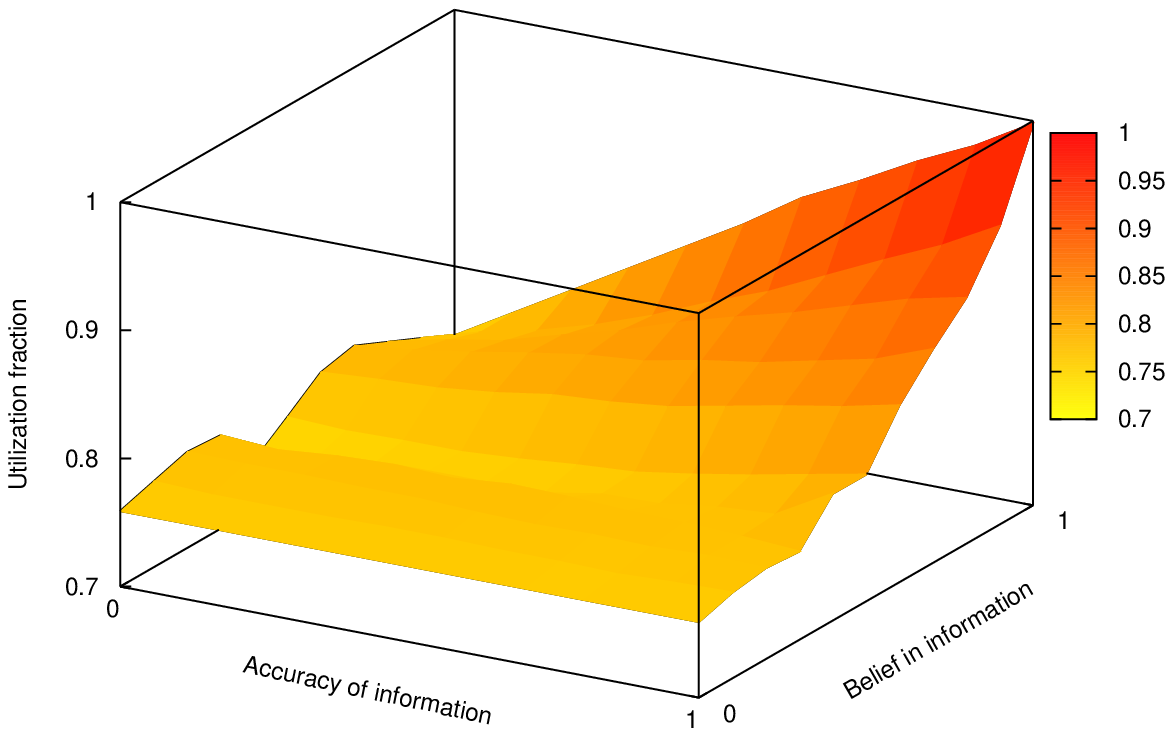} & \includegraphics[width=7cm]{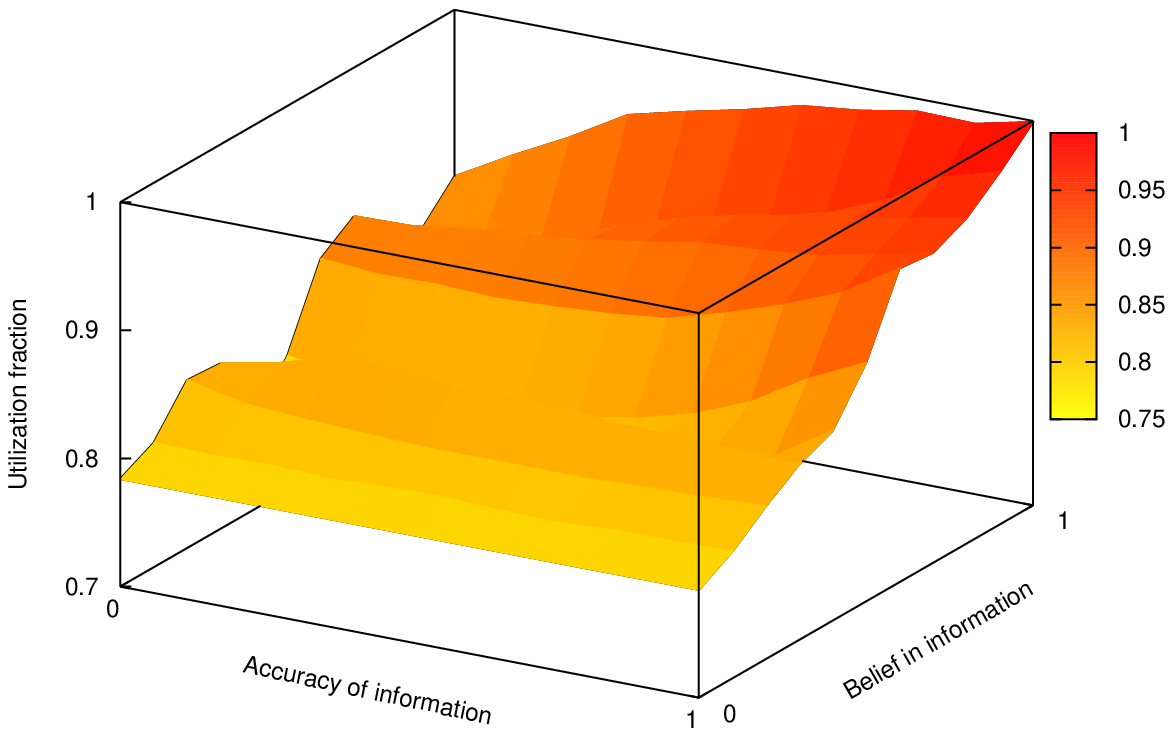}\\
		\bottomrule
	\end{tabular}
	\caption{(Color online) Utilization fractions for various revision protocols.}
	\label{fig:u2}
\end{figure}

\begin{figure}[!tb]	\centering
	\begin{tabular}{cc}
		\includegraphics[width=7cm]{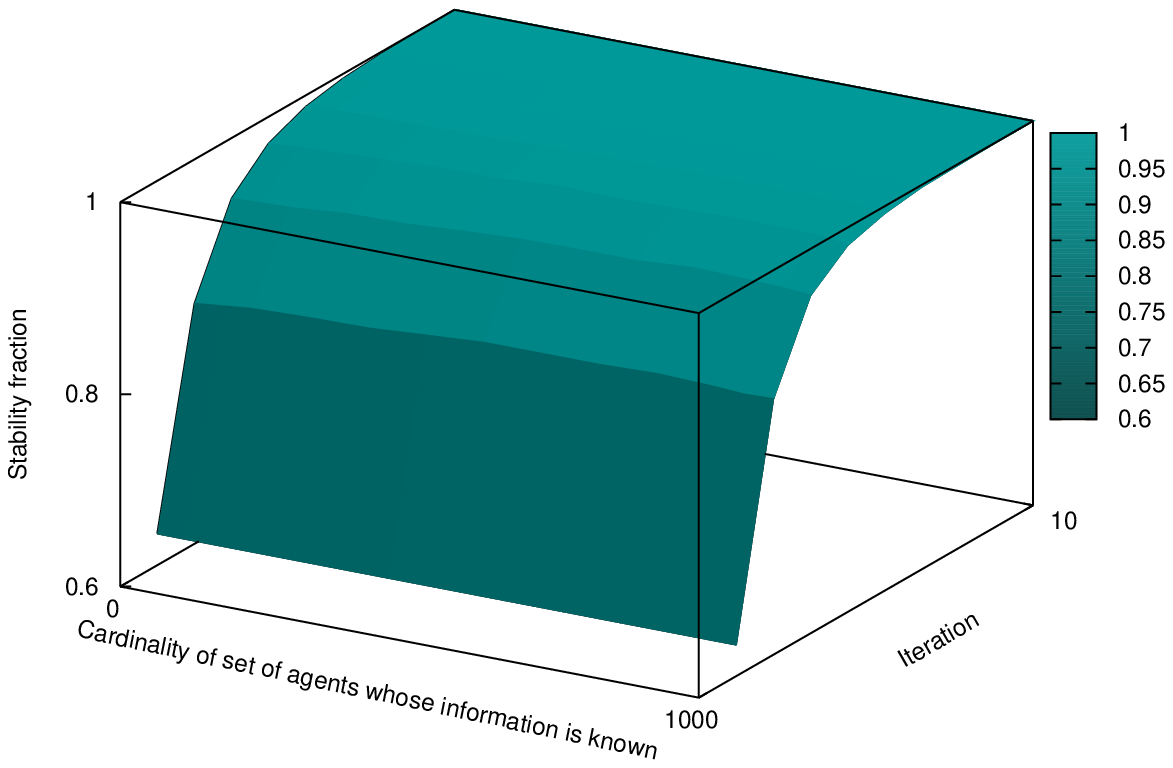} & 
		\includegraphics[width=7cm]{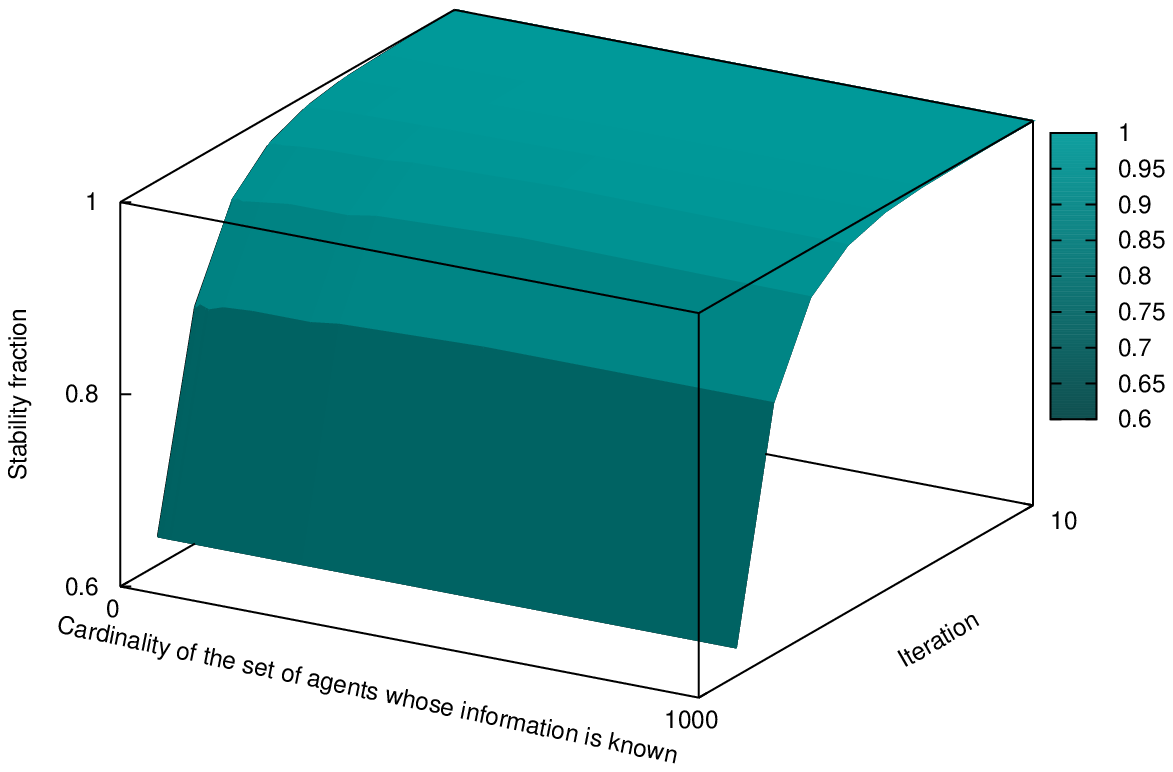}\\
		RP1 & RP2 \\
	\end{tabular}
	
	\begin{tabular}{cc}
		\includegraphics[width=7cm]{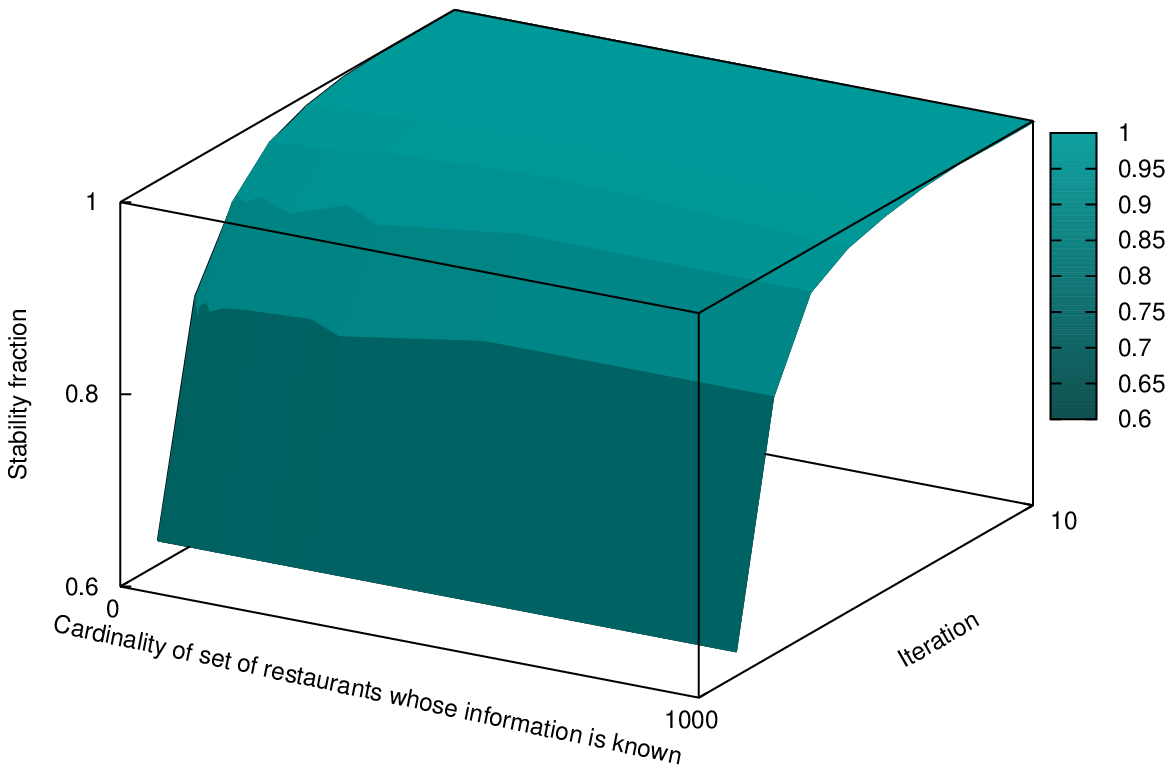} & 
		\includegraphics[width=7cm]{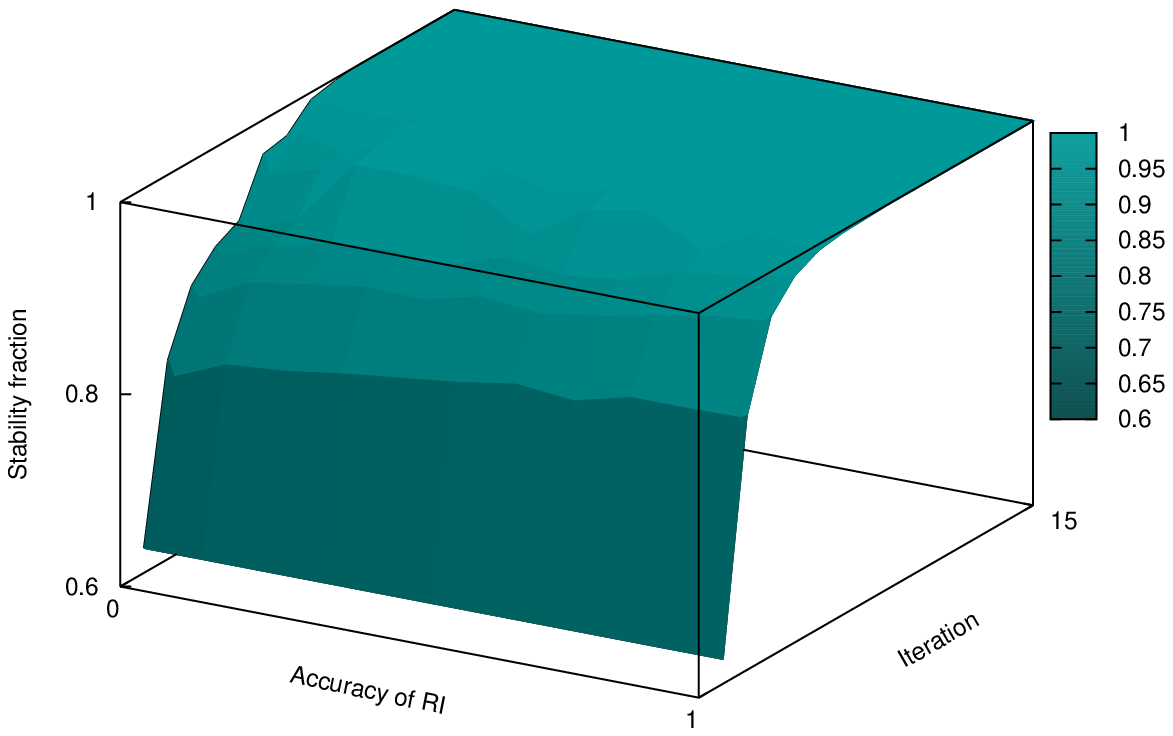} \\
		RP3 & RP4\\
	\end{tabular}
	
	\begin{tabular}{c}
		\includegraphics[width=7cm]{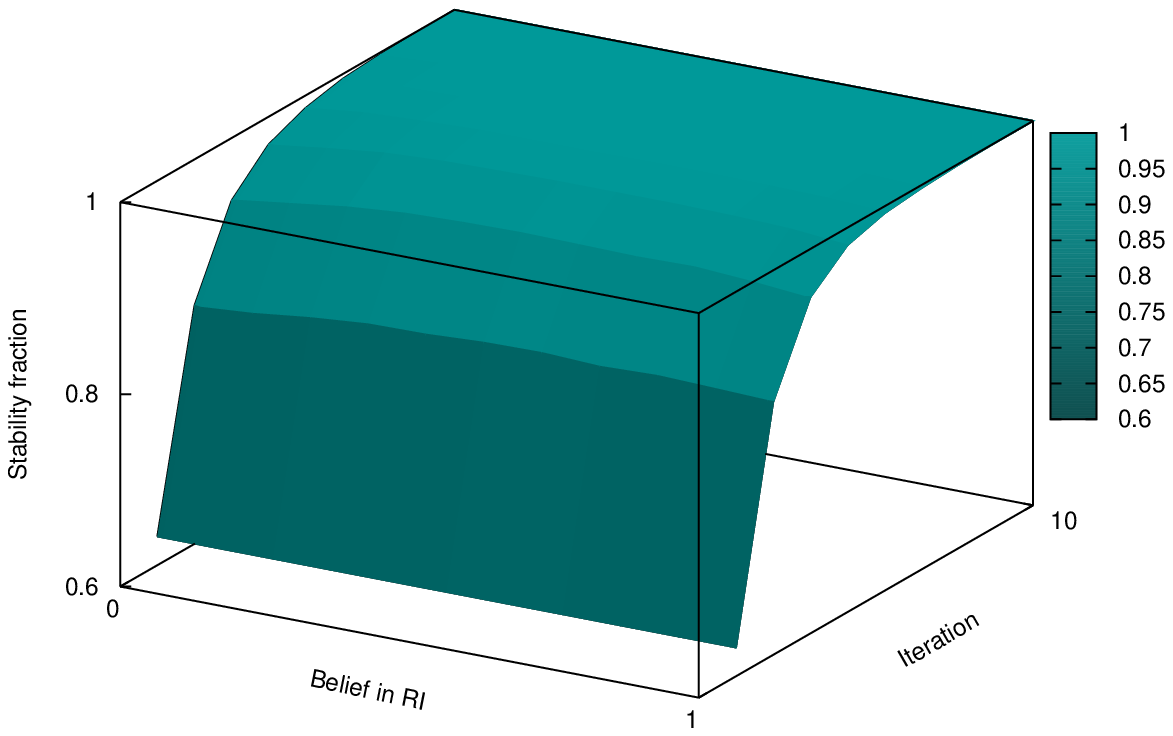}\\
		RP5\\
	\end{tabular}
	\caption{(Color online) Stability fractions for Variant~1 of various revision protocols.}
	\label{fig:s}
\end{figure}

For our simulations we set the value of $N$ to 1000. We track the performance of the two variants when the customers' probability vectors were revised as described in RP1 through RP5 through 20 periods. When customers' probability vectors were revised as described in RP6, we found out the utilization fraction values ultimately reached for different accuracy and belief levels and plotted them. Figures~\ref{fig:u1} and~\ref{fig:u2} show the variations of the utilization fraction values attained by both variants of all six revision protocols. Figure~\ref{fig:s} shows the variations of stability fractions of the first variant when the customers' probability vectors were revised as described in RP1 through RP5 through 20 periods. As mentioned earlier, stability ratios are equal to utilization fractions for all periods in the second variant.

\afterpage{\clearpage}
The plots in Fig.~\ref{fig:s} show that the stability fractions for the five revision protocols (RP1 through RP5) reach a value of 1 after at most 20 periods. This means that for Variant~1 of these five revision protocols the utilization fraction values will not increase from the values seen at the end of 20 periods (see the Variant~1 column in Figs.~\ref{fig:u1} and~\ref{fig:u2}). So for RP1 through RP3, the final value of utilization fractions increase from a value of approximately 0.8 when customers have no information about other customers or restaurants to 1 when customers have complete information. The results from RP6 show interesting results. We see that as the accuracy of information increases the value of the utilization fraction increases, when the belief in the information is high. When the belief in the information is low, the utilization fraction does not change appreciably with the accuracy of information. This is understandable, since in these revision protocols, customers mostly disregard the information provided. We see a strange trend in utilization fraction values when the accuracy of information is low. There the utilization fraction does not increase monotonically with increase in belief in the information. This is probably because belief in worthless information is capable of making customers take incorrect decisions.

In all six revision protocols, we see that the utilization fraction obtained by following Variant~2 of a revision protocol results in utilization fractions that are better than those obtained by Variant~1. This shows that it is better for customers to re-consider their options if restaurants do not serve them. 

\begin{figure}[htb]\centering
	\includegraphics[width=10cm]{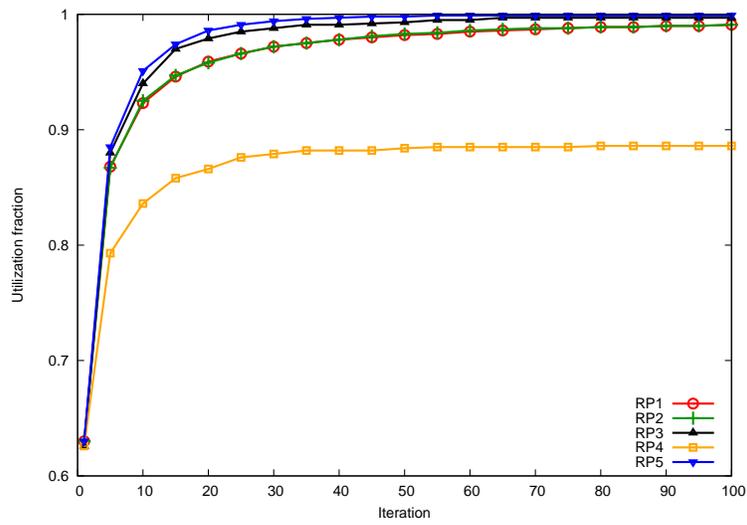}
	\caption{(Color online) Utilization fractions for various revision protocols over 100 iterations.}
	\label{fig:100iter}
\end{figure}

Since the values of stability ratios reach unity for Variant~1 of all revision protocols at the end of 20 iterations, we know that the utilization fractions reached by following those revision protocols at the end of 20 iterations is the best that can be obtained from them. However, it is not clear from our simulations whether this is the revision protocol for Variant~2 of each revision protocol. We therefore performed another simulation using Variant~2 of the first five revision protocols. In this simulation, we simulate the performance of Variant~2 of the first five revision protocols for 100 iterations. For RP1 and RP2 we let customers have information about 5\% of other customers in the previous period. In RP3 we let customers know what happened to 5\% of the restaurants in the previous period. In RP4 we assume that the information about the set of idle restaurants is 5\% accurate, while in RP5 we allow customers to have 5\% belief in the information provided. Fig.~\ref{fig:100iter} reports our findings from this simulation.
From the figure we see that even if the information is limited or even if customers have little belief in the information, the utilization fraction reaches a value close to 1 after a few iterations, cf., RP1, RP2, RP3, and RP5. However, if the information is inaccurate, cf.\ RP4, the value of the utilization fraction can be significantly lower.

\section{Discussion and summary}
\label{sec:summary}
In this paper, we provide some efficient algorithms to solve a particular problem of distributed coordination, namely the KPR problem. The basic problem is to devise strategies with low computational cost for a set of independent agents who are competing for resources, to maximize the rate of resource utilization. Several adaptive strategies have been proposed earlier in the literature, which lead to high utilization rates. Here, we propose revision protocols that depend on reinforcement learning which can lead lead to dramatic improvements in the utilization rate with very few errors using very limited information sets. Hence, such protocols are efficient.

In particular, even with identical agents and identical resources, Pavlovian win-stay, lose-shift strategy (extreme reinforcement learning) leads to dynamic inefficiency as multiple agents can get stuck with one resource. This problem can be solved by assigning a rule that whichever agent chose a particular resource first, will have monopoly over the resource it occupies. However, such a scheme is very inefficient for the agents who lose in the initial rounds and such a scheme is not symmetric across agents anymore. We show that such problems can also be tackled easily by introducing the possibility of switching in case of a loss for all agents. Our numerical results indicate that such solutions with such probability revisions lead to very efficient use of available resources.  
Potential applications of the proposed solutions include solving traffic jams \cite{Tlig2013}, collaborative models of distributed robotic systems \cite{Lerman_01}, client/server computing \cite{Adler1995} among others.

\end{document}